# Title: Electrically Tunable Excitonic Light Emitting Diodes based on Monolayer $WSe_2$ p-n Junctions

**Authors:** Jason S. Ross[1], Philip Klement[2,3], Aaron M. Jones[3], Nirmal J. Ghimire[4,5], Jiaqiang Yan[5,6], D. G. Mandrus[4,5,6], Takashi Taniguchi[7], Kenji Watanabe[7], Kenji Kitamura[7], Wang Yao[8], David H Cobden[2], Xiaodong Xu[1,2*]

Affiliations:

[1]Department of Materials Science and Engineering, University of Washington, Seattle, Washington 98195, USA

[2]Department of Physics, Justus Liebig University, 35392 Giessen, Germany

[3]Department of Physics, University of Washington, Seattle, Washington 98195, USA

[4]Department of Physics and Astronomy, University of Tennessee, Knoxville, Tennessee 37996, USA

[5]Materials Science and Technology Division, Oak Ridge National Laboratory, Oak Ridge, Tennessee, 37831, USA

[6]Department of Materials Science and Engineering, University of Tennessee, Knoxville, Tennessee, 37996, USA

[7]Advanced Materials Laboratory, National Institute for Materials Science, Tsukuba, Ibaraki 305-0044, Japan

[8]Department of Physics and Center of Theoretical and Computational Physics, University of Hong Kong, Hong Kong, China

*Correspondence to: xuxd@uw.edu

**Abstract: Light-emitting diodes are of importance for lighting, displays, optical interconnects, logic and sensors[1–8]. Hence the development of new systems that allow improvements in their efficiency, spectral properties, compactness and integrability could have significant ramifications. Monolayer transition metal dichalcogenides have recently emerged as interesting candidates for optoelectronic applications due to their unique optical properties[9–16]. Electroluminescence has already been observed from monolayer $MoS_2$ devices[17,18]. However, the electroluminescence efficiency was low and the linewidth broad due both to the poor optical quality of $MoS_2$ and to ineffective contacts. Here, we report electroluminescence from lateral p-n junctions in monolayer $WSe_2$ induced electrostatically using a thin boron nitride support as a dielectric layer with multiple metal gates beneath. This structure allows effective injection of electrons and holes, and combined with the high optical quality of $WSe_2$ it yields bright electroluminescence with 1000 times smaller injection current and 10 times smaller linewidth than in $MoS_2$[17,18]. Furthermore, by increasing the injection bias we can tune the electroluminescence between regimes of impurity-bound, charged, and neutral excitons. This system has the required ingredients for new kinds of**

**optoelectronic devices such as spin- and valley-polarized light-emitting diodes, on-chip lasers, and two-dimensional electro-optic modulators.**

**Main Text**

Few-layer group-VIB transition metal dichalcogenides (TMDs) represent a class of semiconductors in the two-dimensional (2D) limit[9,10,19]. Due to their large carrier effective mass and the reduced screening in 2D, electron-hole interactions are much stronger than in conventional semiconductors. This leads to large binding energies for both charged and neutral excitons which as a result are spectrally sharp, robust, and amenable to electrical manipulation[16,20,21]. In addition, the large spin-orbit coupling[22] and the acentric structure of TMDs provides a connection between spin and valley degrees of freedom[14], light polarization[11,13,15,16], and magnetic and electric fields[23] that can be exploited for new kinds of device operation.

Although in bulk TMDs the band gap is indirect, in the limit of a single monolayer it becomes direct[9,10], fulfilling the most basic requirement for efficient light emission. Indeed, electroluminescence (EL) has already been reported from monolayer $MoS_2$ field-effect transistors (FETs), occurring near the Schottky contact with a metal[17] or with highly doped silicon[18]. However, the efficiency and spectral quality was much lower than has been demonstrated for other nanoscale light emitters such as carbon nanotubes[7], for two reasons. First, efficient EL requires effective injection of both electrons and holes into the active region, which should therefore be within a p-n junction. Second, $MoS_2$ is known to have poorer optical quality than other group VIB TMDs, possibly due to impurities. It has been previously shown that in contrast monolayer $WSe_2$ has excellent optical properties[16,24]. Here, we demonstrate that combining monolayer $WSe_2$ with a p-n junction architecture using electrostatic doping produces efficient and electrically tunable excitonic light-emitting diodes (LEDs).

An optical image and a schematic of a device, made by a combination of electron-beam lithography and transfer of exfoliated sheets, are shown Figs. 1a and 1b. A monolayer $WSe_2$ sheet sits on a sheet of hexagonal boron nitride (BN), typically 10 nm thick, which acts both as a smooth, disorder-free substrate to minimize non-radiative energy relaxation pathways and as a high quality gate dielectric. Applying voltages to the two 7 nm palladium gate electrodes beneath the BN can create two separate electrostatically doped regions in the $WSe_2$ separated by a 300 nm wide undoped strip. Gold/vanadium (60/6 nm) source and drain contacts are evaporated on top.

Importantly, they overlap the gates in order to reduce the Schottky barriers at the contacts. For electrical transport measurements a dc bias $V$ is applied to one contact (the source) and the current $I$ from the other (the drain) is measured by a virtual-earth current preamplifier. The silicon substrate is grounded.

We start by showing that a p-n junction can be created electrostatically. First we set the gate voltages $V_{g1}$ and $V_{g2}$ to the same value. Figure 1c shows the current produced by a bias $V = 0.5$ V as the gate voltage $V_{g1} = V_{g2}$ is swept at 60 K (room temperature measurements are shown in Supplementary Figure 1). The current increases rapidly for gate voltages > +6.5 V (electron doping) and < -6.5 V (hole doping), demonstrating ambipolar operation. In the inset the red $I$-$V$ curve, taken at $V_{g1} = V_{g2} = +8.0$ V, is almost symmetric as expected for both gated regions being equally electron-doped. The nonlinearity near zero bias can be associated with the undoped gap between the gates and residual Schottky barriers at the contacts. In contrast, the blue $I$-$V$ curve, taken with $V_{g1} = +8.0$ V and $V_{g2} = -8.0$ V, shows the strong rectification behavior expected for a p-n junction.

The p-n junction can be investigated in detail by scanning optical measurements[25,26]. Fig. 2a is a microscope image of a device and Fig. 2b is a corresponding scanning photocurrent image, measured with zero bias at 100 K using a 10 µW diffraction-limited 660 nm laser spot scanned over the sample. We see a large photocurrent signal localized between the gates, with a peak magnitude of 5 nA. Taking into account the 1% absorption of $WSe_2$ monolayers at 660 nm[27], the internal quantum efficiency reaches a maximum of about 5%. Such a photocurrent is the natural result of the junction functioning as a photodiode, with photogenerated carriers separated by a strong depletion field concentrated in the undoped gap. The sensitivity of the photodiode can be tuned over a wide range by varying the gate voltages and bias (see Supplemental Figure 2).

Fig. 2c is a corresponding map of the integrated photoluminescence (PL) intensity, which indicates the high optical quality of the whole $WSe_2$ sheet and shows that the luminescence is not substantially quenched by the underlying gates. More revealingly, Fig. 2d shows a colour map of the peak PL photon energy, exhibiting two distinct regions clearly correlated with the expected n-doped (blue) and p-doped (red) parts of the $WSe_2$ above the gates. The reason for this is made clear in Fig. 2e, which shows PL spectra taken at different positions. The detailed origin of these spectral features has been established previously[16]. Above the gate held at $V_{g1}$ = +8.0 V (blue trace) the negatively charged X⁻ trion (two electrons and one hole) dominates, implying an excess of electrons. Above the other gate, held at $V_{g2}$ = -8.0 V (red trace), the higher-energy positively charged X⁺ trion (two holes and one electron) dominates, implying an excess of holes. In the gap between the gates (black trace) the neutral exciton X⁰ peak can also be seen, consistent with no doping in that region. Here the superimposed X⁺ and X⁻ peaks may come from the gated regions, since the laser spot is larger than the gap.

When the device is configured as a p-n junction ($V_{g1}$ = -$V_{g2}$ = 8 V), but not otherwise, we observe bright electroluminescence. Good spectra can be obtained even at room temperature with a current of 5 nA, as illustrated (blue) in Fig. 3a. This current is three orders of magnitude smaller than in MoS₂ FETs[17,18]. In fact, in our best device we observed EL at an injection current as low as 200 pA (See Supplemental Figure 3). To understand the nature of the EL, we superimpose a normalized PL spectrum of undoped monolayer WSe₂ (red). It is known that the PL of WSe₂ is from the recombination of direct-gap excitons; thus the similarity between the EL and PL spectra implies that the injected electrons and holes form excitons before recombining radiatively. This is a natural consequence of the large exciton binding energy due to the strong Coulomb interaction in monolayer TMDs. Fig. 3b shows an image of the total EL intensity (coloured) superimposed on a simple white-light reflection image (grayscale). It is clear that the EL emanates from the entire length of the monolayer junction between the two gates.

At low temperatures the EL spectrum develops interesting structure. Fig. 4a shows a plot of EL intensity as a function of current and photon energy. There are three main spectral features: a narrow higher-energy peak (green arrow), a broad central peak (brown arrow), and a lower-energy peak (black arrow). The shapes and relative intensities of these features change with current. Their origins can be deduced from a comparison with the PL, whose intensity is plotted in Fig. 4b as a function of photon energy and common gate voltage $V_g=V_{g1}=V_{g2}$. Here we see the tuning of the dominant exciton species as the carrier density is changed by gating[16]. The PL feature which is strongest at $V_g$ = 0 is due to X⁰ recombination. It has a similar width to, and is at the same position (1.69 eV) as, the narrow EL peak, which we therefore identify as the X⁰ emission. The dominant PL feature at $V_g$ > 0, which shifts from 1.663 eV to 1.625 eV as $V_g$ increases from 0 to +8 V, is due to X⁻ (of which there are multiple species[16]). The broad EL feature occurs in the same range of energies, implying that it is dominated by $X^-$. The dominant feature in the PL at $V_g < 0$ is due to $X^+$. This aligns with the high-energy shoulder (grey arrow) on the broad EL feature at about 1.670 eV. There is also a band of emission from impurity-bound excitons ($X^I$) in the PL which matches the lower-energy EL feature centered at 1.59 eV.

The finding that $X^-$ dominates the EL in Fig. 4a is consistent with the observation that $X^-$ has much stronger PL than the other exciton species (Fig. 4b). The shifts of the trion peaks in the PL with $V_g$ imply that the trion binding energy depends on perpendicular electric field. Hence the fact that the width of the broad peak in the EL matches the full range of $X^-$ energies in the PL is explained by the variation of the perpendicular electric field across the junction. On the other hand the $X^0$ peak shifts very little with $V_g$ in the PL and hence is insensitive to field; thus the $X^0$ EL peak is also sharp. The $X^0$ EL linewidth is found to be as narrow as 5 meV, which is an order of magnitude smaller than for EL from $MoS_2$[1,7].

In Fig. 4c we show the EL spectra at selected current values, illustrating sequential population of the excitonic states, which could be due to current pumping or to changes in the electric field at the junction under different source-drain biases. At the lower current (24 nA) only the excitons with lowest energy ($X^I$ and $X^-$) are seen. At a higher current (27 nA) the $X^+$ shoulder appears, and at 31 nA an $X^0$ peak is also present. At the highest current (33 nA) we illustrate how the spectrum can be decomposed into four Gaussian peaks. It is also apparent that the relative strength of the $X^I$ peak decreases as the current increases.

The above observations reflect complex exciton dynamics which are not yet fully understood. Time-resolved PL measurements have shown that the lifetimes (for radiative and nonradiative processes combined) of the free excitons and the impurity-bound excitons are about 5 ps and 100 ps respectively[28–31] in monolayer $MoS_2$, which we expect to be similar to $WSe_2$. The electron-hole pair injection rate is $\frac{I}{e}$ ($e$ is electron charge), which is one pair per 5 ps at 32 nA, comparable to the lifetime of a free exciton but much shorter than that of $X^I$. Therefore, there could be only of order one free exciton but many impurity-bound excitons present in the junction. We speculate that that the presence of multiple $X^I$ combined with the strong Coulomb interactions enhances non-radiative recombination, which limits the $X^I$ emission at higher currents. Alternatively, the saturation of the

X[1] peak could also be due to the filling of impurity states as observed in standard PL and photocurrent experiments.

We measure the total photon emission rate at the largest applied current of 35 nA to be about 16 million sec$^{-1}$ (Supplemental Note 1). This is 10 times larger, for 1000 times smaller current, than reported for MoS$_2$ devices[17,18]. It corresponds to 1 photon per $10^4$ injected electron-hole pairs. We expect that the overall device efficiency could be improved by increasing the injection current for a given voltage by reducing contact resistance, by improving the WSe$_2$ crystal quality, and by employing improved membrane transfer techniques.

Finally we mention an important implication of these results. It has been conclusively demonstrated using polarization-resolved PL that the excitons in monolayer WSe$_2$ are formed in the ±K valleys[16]. The excellent match of the EL with the PL thus proves that the EL also comes from such valley excitons. In the experiments described here, the injected electrons and holes populate both valleys equally as sketched in Fig. 4d, forming excitons in both valleys and thus producing unpolarized light, as we have checked (Supplementary Figure 4). As a result of the spin-valley locking in monolayer TMDs, where the +K and -K band edges have opposite spin[14], we expect that using ferromagnetic contacts to obtain spin-polarized injection will allow the creation of spin- and valley-LEDs with controllably polarized emission[32].

During the review process, we became aware of two related pieces of work which were posted on arXiv[33,34].

**Acknowledgments:** This work is mainly supported by the US DoE, BES, Materials Sciences and Engineering Division (DE-SC0008145). NG, JY, DM are supported by US DoE, BES, Materials Sciences and Engineering Division. WY is supported by the Research Grant Council of Hong Kong (HKU705513P), the University Grant Council (AoE/P-04/08) of the government of Hong Kong, and the Croucher Foundation under the Croucher Innovation Award. D.C. is supported by US DoE, BES, Materials Sciences and Engineering Division (DE-SC0002197). Device fabrication was performed at the Washington Nanofabrication Facility and NSF-funded Nanotech User Facility.

**Author Contributions:** X.X. conceived the experiments. J.S.R. fabricated the devices and performed the measurements, assisted by P.K. and A.M.J.. J.S.R. and X.X. performed data analysis, with input from D.C. and W.Y.. N.G., J.Y. and D.G.M. synthesized and performed bulk characterization measurements on the WSe$_2$ crystals. T.T., K.W. and K.K. provided boron nitride crystals. X.X, D.C., J.S.R and W.Y. wrote the paper. All authors discussed the results.

**Competing Final Interests**

The authors declare no competing financial interests.

**Figures.**

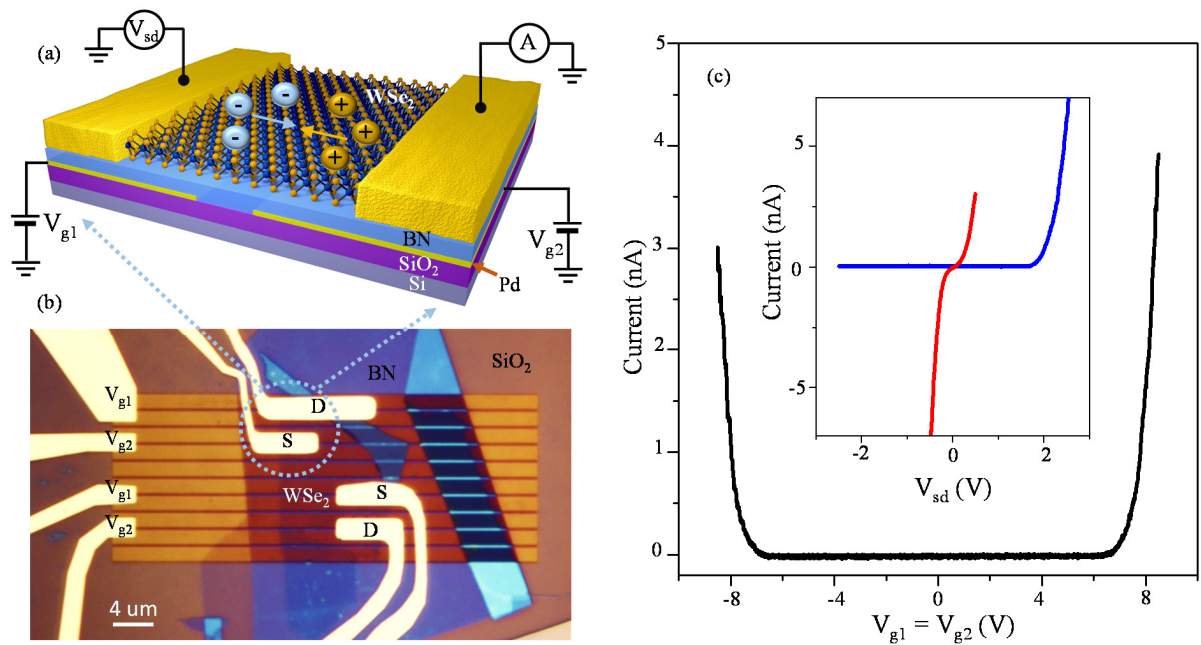

**Figure 1 | Monolayer WSe₂ p-n junctions. a**, Cartoon and **b**, optical micrograph of multiple monolayer WSe$_2$ p-n junction devices with palladium back gates ($V_{g1}$ and $V_{g2}$) and source (S) and drain (D) contacts. The source-drain voltage ($V_{sd}$) is applied to one contact and the current (A) is read out of the other. During electroluminescence in the WSe$_2$, electrons (blue) and holes (yellow) move towards each other (arrows) and recombine. The back gates are separated from the WSe$_2$ by hexagonal boron nitride. The device sits on a layer of silicon dioxide on a silicon substrate. **c**, Main panel: variation of current with both bottom gate voltages set equal ($V_{g1}=V_{g2}=V_g$) for one junction at bias $V_{sd}$ = 500 mV, showing ambipolar behavior. Inset: $I$-$V_{sd}$ characteristics when the two gate voltages are set to equal ($V_{g1}=V_{g2}=8$ V, red) and opposite ($V_{g1}=-8$ V, $V_{g2}=8$ V, blue) values above the injection threshold.

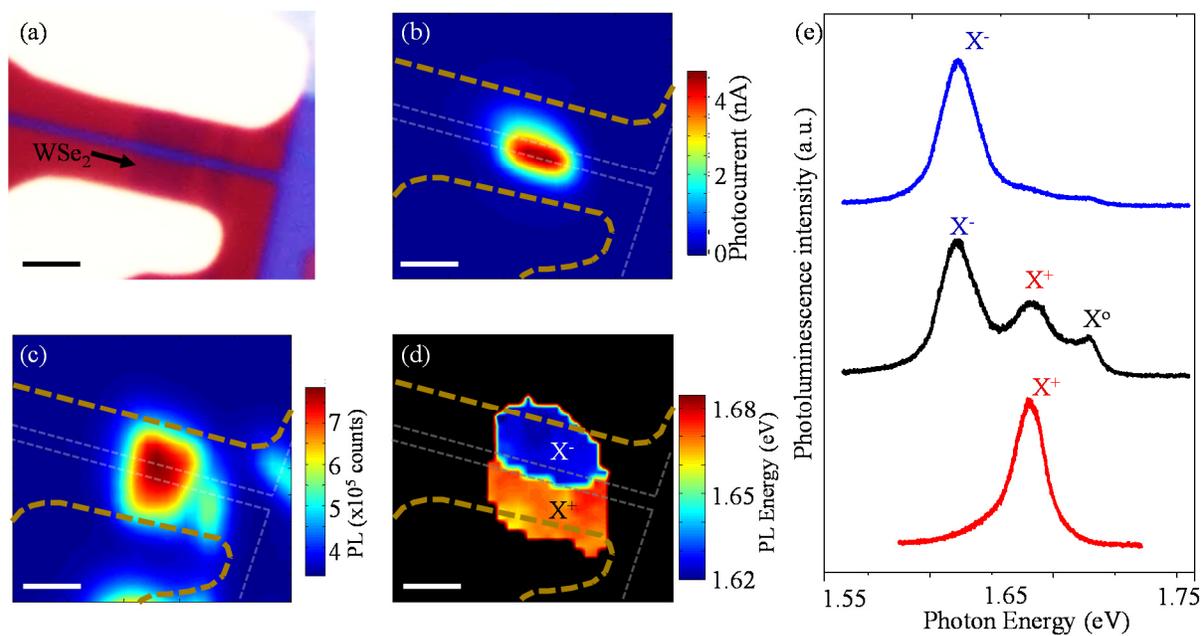

**Figure 2 | Photoresponse of monolayer p-n junction at 100 K. a**, Microscope image of a monolayer p-n junction device. The source and drain contacts are white, the two bottom gates are red, and the boron nitride is blue. **b**, Corresponding scanning photocurrent image showing pronounced photocurrent generation localized to the junction. The thicker dashed lines outline the source and drain contacts while the thinner dashed lines outline the back gates. **c**, Integrated photoluminescence map. **d**, Peak photoluminescence energy map showing p and n regions as a result of the different energies of oppositely charged excitons. **e**, Top to bottom: selected spectra from n-doped region (blue), junction (black), and p-doped region (red). All maps taken with $V_{g1}$ = +8 V, $V_{g2}$ = -8 V. Scale bars: 4 μm.

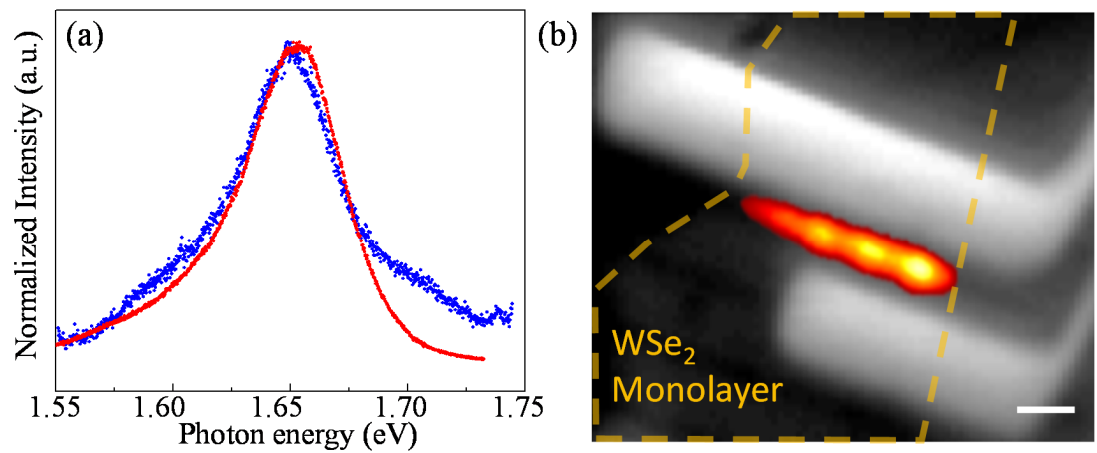

**Figure 3 | Photoluminescence (PL) and electroluminescence (EL)**. **a**, At 300 K, the EL spectrum (blue) generated by a current of 5 nA closely resembles the PL spectrum (red). **b**, EL image (red) superimposed on device image (grayscale). The orange dashed lines outline the WSe$_2$ monolayer. Scale bar: 2 μm.

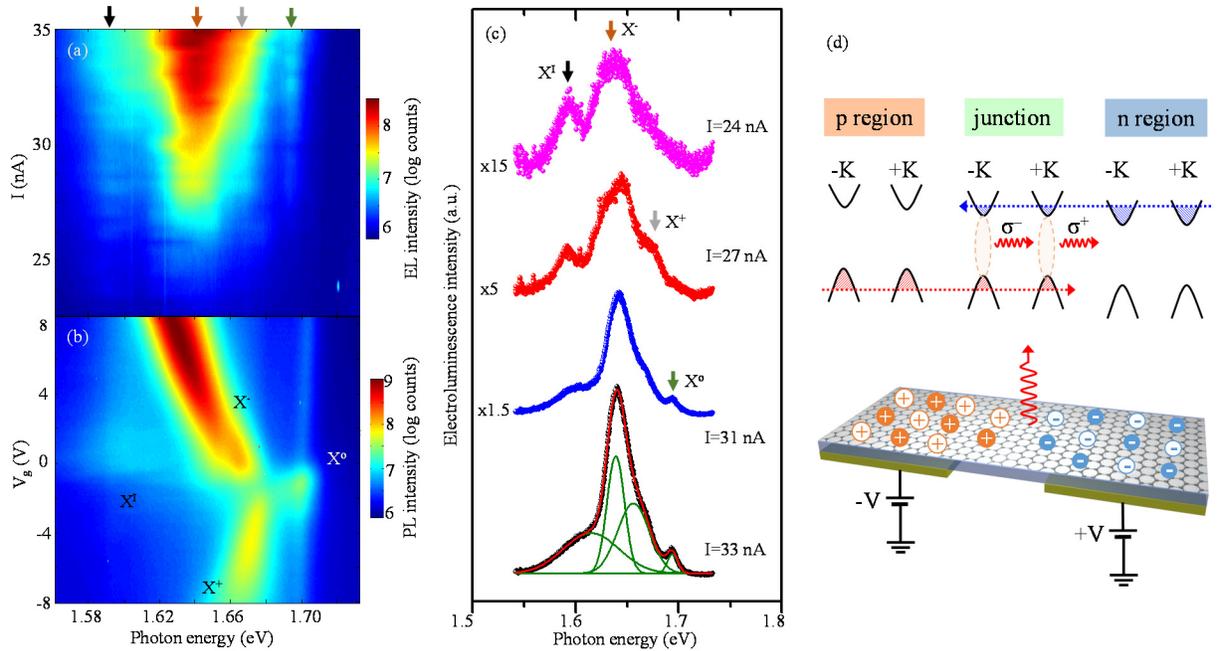

**Figure 4 | Tuning valley-exciton electroluminescence at 60 K. a**, Electroluminescence (EL) intensity plot as function of bias current and photon energy. From left to right, the arrows indicate the impurity-bound exciton ($X^I$), the charged excitons ($X^-$ then $X^+$), and the neutral exciton ($X^o$). **b**, Plot of Photoluminescence (PL) intensity as a function of photon energy and gate voltage $V_g=V_{g1}=V_{g2}$. **c**, Selected EL spectra at different bias currents. As the current increases, we observe EL tuning from $X^I$ through $X^-$ and $X^+$ and finally $X^o$. The bottom spectrum is fit by four Gaussian lineshapes, one for each exciton species. **d**, Band diagram and device schematic showing EL generation from valley excitons. Filled and empty circles indicate carriers in the +K and -K valleys. Both valleys are shown to be populated leading to EL that has both right ($\sigma^+$) and left ($\sigma^-$) circular polarization.

# Title: Electrically Tunable Excitonic Light Emitting Diodes based on Monolayer WSe$_2$ p-n Junctions

**Authors:** Jason S. Ross[1], Philip Klement[2,3], Aaron M. Jones[3], Nirmal J. Ghimire[4,5], Jiaqiang Yan[5,6], D. G. Mandrus[4,5,6], Takashi Taniguchi[7], Kenji Watanabe[7], Kenji Kitamura[7], Wang Yao[8], David H Cobden[2], Xiaodong Xu[1,2*]

**Supplementary Figures**

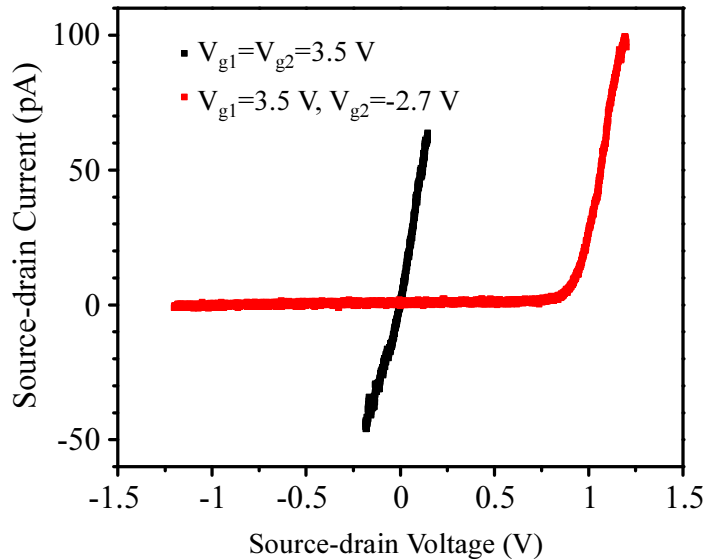

**Figure S1| Electrical properties of monolayer WSe$_2$ p-n junction at room temperature.** This device had a thinner layer of boron nitride (8 nm) than the device featured in the main text. Therefore lower gate voltages were required to modulate the carrier density.

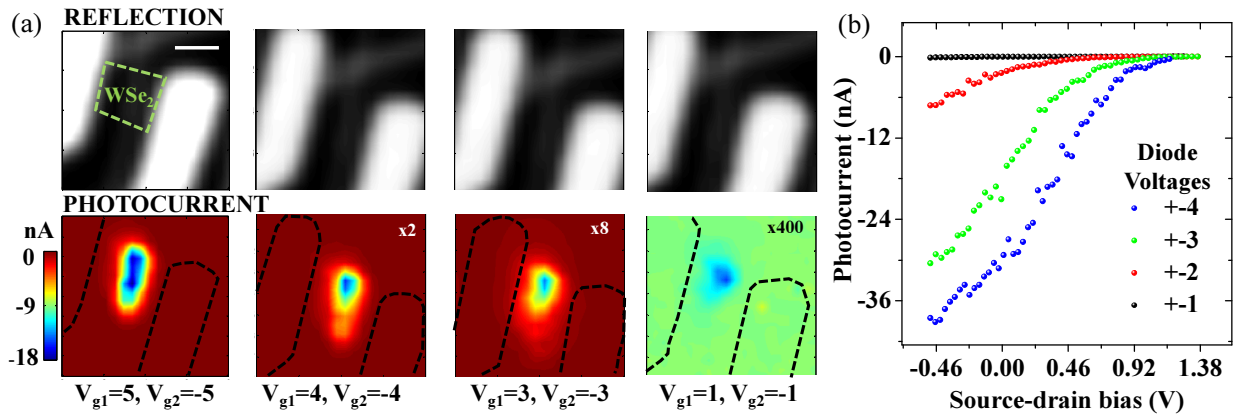

**Figure S2| Photocurrent measurements**. **a**, Scanning-laser (10 µW at 532 nm) reflection (top row) and corresponding photocurrent (PC, bottom row) images at different p-n junction fields as controlled by the opposite split-gate voltages. The PC is generated at gate voltages as low as ±1 V. The location and shape of the PC spot, especially at high gate voltages, demonstrates that the p-n junction in the gap between the split gates governs the photoresponse, as opposed to Schottky barriers at the contacts. **b**, Photocurrent as a function of bias showing high sensitivity.

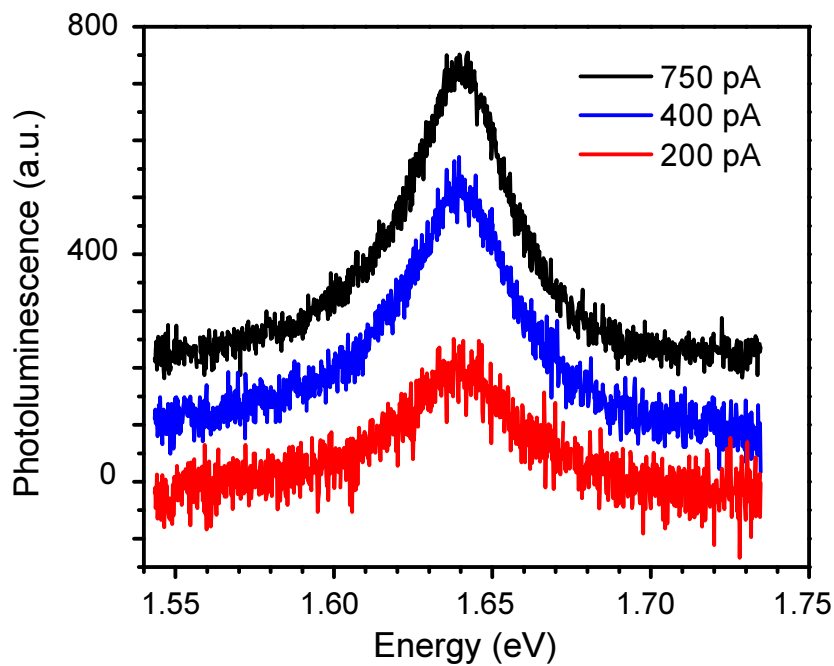

**Figure S3| Minimum Electroluminescence Detection at Room Temperature.** Background subtracted and artificially offset spectra of the electroluminescence at various injection currents. Electroluminescence is resolvable for injections currents as low as 200 pA.

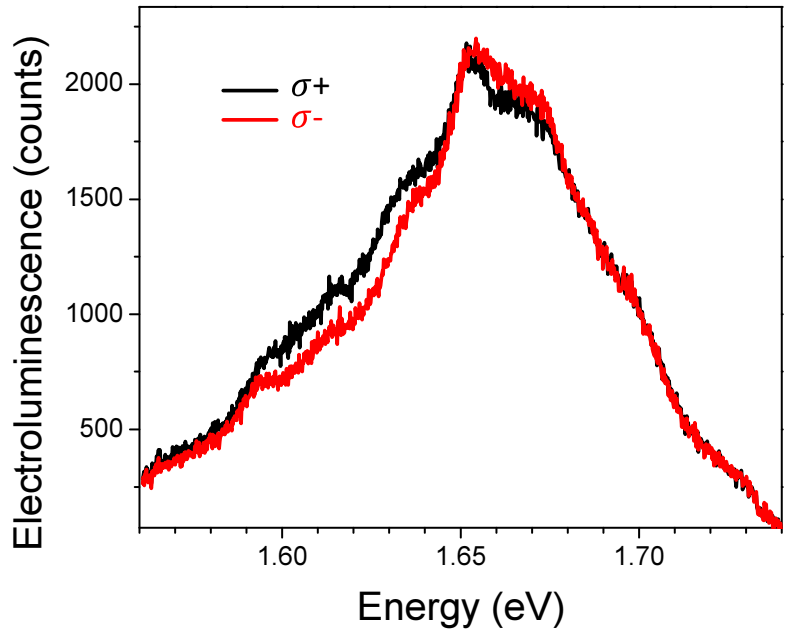

**Figure S4| Electroluminescence Polarization.** Electroluminescence (EL) under left ($\sigma$-) and right ($\sigma$+) polarization detection showing no appreciable polarization. The difference in left and right circularly polarized EL shows inconsistent fluctuations and never exceeds 5%, the systematic error of this measurement due to optics. Currently, we attribute any perceived polarization to the time-dependent fluctuations in the EL.

**Supplementary Note 1: Calibration of Collection Efficiency.**

For electroluminescence (EL) measurement, the signal is collected by a 40X, NA 0.6 objective, passed through the 685 nm dichroic beam splitter and sent into the spectrometer setup. The efficiency of the system is $\alpha = \alpha_{obj}\alpha_{NA}\alpha_{spec}\beta_{spec}$, where $\alpha_{obj}$ is the transmission efficiency of the object leg, $\alpha_{NA}$ is the collection efficiency of the objective itself limited by numerical aperture, $\alpha_{spec}$ is the collection efficiency of the spectrometer leg, and $\beta_{spec}$ is the ratio of efficiencies of the spectrometer components (grating and CCD) at 660 nm (our calibration laser, see below) to those in the EL spectral range. The CCD and grating specification sheets were used to estimate $\beta_{spec}$. A Thorlabs S120VC power meter and a 660 nm diode laser were used to calibrate $\alpha_{obj}$ and $\alpha_{spec}$. Finally, the objective collection efficiency is given by the integral over the solid angle defined by the numerical aperture[1,2]:

$$\alpha_{NA} = \frac{3}{4\pi} \int_0^{2\pi} d\phi \int_0^{\arcsin(NA/n)} d\theta \sin\theta = \frac{3}{2}\left(1 - \sqrt{1 - \left(\frac{NA}{n}\right)^2}\right).$$

In our case, $n = 1$ since the WSe$_2$ is not covered with any material. The total photon emission rate from the junction is then $G = G_{\text{measured}}/\alpha$. The rate of electron and hole injection is $I/e$, and hence the efficiency (number of photons produced per injected electron) is $(e/I)G$.